\documentclass{ws-procs9x6}

\newcommand{\vk}{\mbox {\boldmath $k$\unboldmath}}
\newcommand{\vq}{\mbox {\boldmath $q$\unboldmath}}
\def\nn{\nonumber}

\begin{document}

\title{Role of the baryon resonances in the $\eta$ and $K^+$ photoproduction processes
on the proton}

\author{B. SAGHAI}

\address{DAPNIA, DSM, CEA/Saclay, 91191 Gif-sur-Yvette, France\\
E-mail: bsaghai@cea.fr}


\maketitle
\abstracts{
Very recent $\eta$ and $K^+$ photoproduction data on the proton from 
threshold up to $E_\gamma ^{lab}$ $\approx$ 2.6 GeV are interpreted within 
a chiral constituent quark formalism, which embodies all known nucleon 
and hyperon resonances.
Possible contributions from an additional $S_{11}$ resonance are presented.}
\section{Introduction}
%
Recent experimental and theoretical investigations on the photoproduction 
of mesons\cite{GWU}
are providing us with new insights into the baryon spectroscopy. The present
manuscript is devoted to the interpretation of the processes
\begin{eqnarray}\label{eq:Reac}
\gamma~p \to \eta~p~, K^+ \Lambda.
\end{eqnarray}   

Those reactions have been widely studied {\it via} Effective Lagrangian Approaches 
(ELA) for both $\eta$-meson\cite{RPI,Tiator,Tr,Az} and 
kaon\cite{Cot,Ben,AS,You,Bel} channels. 
Such studies, often based on the Feynman 
diagrammatic technique and embodying {\it s-},{\it u-}, and {\it t-}channel exchanges,
have produced various models differing mainly in their content of baryon resonances.
The number of exchanged particles dealt with in those isobaric models is 
limited\cite{BS1} by 
the number of related free parameters, which increases rapidly from 1 to 5 per 
resonance\cite{OS}, in including resonances with spin $\ge$3/2. 
Given the large number of known resonances (see 
Table 1), such a shortcoming renders those phenomenological approches 
inappropraite\cite{BS2}
in the search for new baryon resonances predicted by various QCD-inspired 
formalisms\cite{Review}. The latter topic is of special interest in the 
present work.
%
%
\begin{table}[t]
\tbl{Baryon resonances from PDG [13], with mass 
$M_{N^*}\leq$ 2.5 GeV. Notations are $L_{2I~2J}(mass)$ and 
$L_{I~2J}(mass)$ for $N^*$ and $Y^*$, respectively.}
{\footnotesize
\begin{tabular}{cll}
\hline
 Baryon &  Three \& four star resonances &  One \& two star resonances \\
\hline
  & $S_{11}(1535)$, $S_{11}(1650)$, & $S_{11}(2090)$, \\
 & $P_{11}(1440)$, $P_{11}(1710)$, $P_{13}(1720)$,
 & $P_{11}(2100)$, $P_{13}(1900)$, \\
{\large $N^*$} & $D_{13}(1520)$, $D_{13}(1700)$, $D_{15}(1675)$, 
 & $D_{13}(2080)$, $D_{15}(2200)$, \\
 & $F_{15}(1680)$,  
 & $F_{15}(2000)$, $F_{17}(1990)$, \\
 & $G_{17}(2190)$, $G_{19}(2250)$,  
 &   \\
 & $H_{19}(2220)$, 
 &   \\ [10pt]
 & $S_{01}(1405)$, $S_{01}(1670)$, $S_{01}(1800)$,  
 &  \\
 &  $P_{01}(1600)$, $P_{01}(1810)$, $P_{03}(1890)$, 
 & \\
{\large $\Lambda^*$} & $D_{03}(1520)$, $D_{03}(1690)$, $D_{05}(1830)$,
 & $D_{03}(2325)$, \\
 & $F_{05}(1820)$, $F_{05}(2110)$,  
 & $F_{07}(2020)$,  \\
 & $G_{07}(2100)$, 
 &   \\
 & $H_{09}(2350)$,
 &   \\ [10pt]
 & $S_{11}(1750)$, 
 & $S_{11}(1620)$, $S_{11}(2000)$, \\
 & $P_{11}(1660)$, $P_{11}(1880)$, $P_{13}(1385)$, 
 & $P_{11}(1770)$, $P_{11}(1880)$,  \\ 
{\large $\Sigma^*$} && $P_{13}(1840)$, $P_{13}(2080)$, \\ 
%
 & $D_{13}(1670)$, $D_{13}(1940)$, $D_{15}(1775)$,
 & $D_{13}(1580)$, \\
 & $F_{15}(1915)$, $F_{17}(2030)$. 
 & $F_{15}(2070)$,   \\
 &  
 & $G_{17}(2100)$.  \\
\hline
\end{tabular}\label{table1} }
\vspace*{-13pt}
\end{table}
%

The content of our chiral constituent quark approach, based 
on the broken $SU(6)\otimes O(3)$ symmetry, is outlined in the next Section.
Comparisons between our models and data are reported in Section 3 and concluding
remarks are given in Section 4.
%
%
\section{Theoretical Frame}
The starting point of the meson photoproduction in the chiral quark model is the low 
energy QCD Lagrangian\cite{MANOHAR}
\begin{eqnarray}\label{eq:Lagrangian}
{\mathcal L}={\bar \psi} \left [ \gamma_{\mu} (i\partial^{\mu}+ V^\mu+\gamma_5
A^\mu)-m\right ] \psi + \dots
\end{eqnarray}
where $\psi$ is the quark field  in the $SU(3)$ symmetry,
$ V^\mu=(\xi^\dagger\partial_\mu\xi+\xi\partial_\mu\xi^\dagger)/2$ 
and 
$A^\mu=i(\xi^\dagger \partial_{\mu} \xi -\xi\partial_{\mu} \xi^\dagger)/2$ 
are the vector and axial currents, respectively, with $\xi=e^{i \Pi f}$. 
$f$ is a decay constant and the field $\Pi$ is a $3\otimes 3$ matrix,
\begin{equation}\label{eq:Pi}
\Pi=\left| \begin{array}{ccc} \frac 1{\sqrt {2}} \pi^\circ+\frac 1{\sqrt{6}}\eta 
& \pi^+ & K^+ \\ \pi^- & -\frac 1{\sqrt {2}}\pi^\circ+\frac 1{\sqrt {6}}\eta & 
K^\circ \\ K^- & \bar {K}^\circ &-\sqrt{\frac 23}\eta \end{array}\right|,
\end{equation}
in which the pseudoscalar mesons, $\pi$, $K$, and $\eta$, are treated
as Goldstone bosons so that the Lagrangian in Eq.~(\ref{eq:Lagrangian}) 
is invariant under the chiral transformation.  
Therefore, there are four components for the photoproduction of
pseudoscalar mesons based on the QCD Lagrangian,
\begin{eqnarray}\label{eq:Mfi}
{\mathcal M}_{fi}&=&\langle N_f| H_{m,e}|N_i \rangle + 
\sum_j\bigg \{ \frac {\langle N_f|H_m |N_j\rangle 
\langle N_j |H_{e}|N_i\rangle }{E_i+\omega-E_j}+ \nn \\
&& \frac {\langle N_f|H_{e}|N_j\rangle \langle N_j|H_m
|N_i\rangle }{E_i-\omega_m-E_j}\bigg \}+{\mathcal M}_T,
\end{eqnarray}
where $N_i(N_f)$ is the initial (final) state of the nucleon, and 
$\omega (\omega_{m})$ represents the energy of incoming (outgoing) 
photons (mesons). 

The pseudovector and electromagnetic couplings at the tree level are given respectively
by the following standard expressions:
\begin{eqnarray}
H_m~=~\sum_j \frac 1{f_m} {\bar \psi}_j\gamma_\mu^j\gamma_5^j \psi_j
\partial^{\mu}\phi_m ~;~
H_e~=~-\sum_j e_j \gamma^j_\mu A^\mu ({\bf k}, {\bf r}).\label{eq:He}
\end{eqnarray}
The first term in Eq.~(\ref{eq:Mfi}) is a seagull term.
The second and third terms correspond to the {\it s-} and {\it u-}channels,
respectively. 
The last term is the {\it t-}channel contribution and is 
excluded here due to the duality hypothesis\cite{LS-2}.

The contributions from  the {\it s-}channel resonances to the transition 
matrix elements can be written as
\begin{eqnarray}\label{eq:MR}
{\mathcal M}_{N^*}=\frac {2M_{N^*}}{s-M_{N^*}(M_{N^*}-i\Gamma(q))}
e^{-\frac {{k}^2+{q}^2}{6\alpha^2_{ho}}}{\mathcal A}_{N^*},
\end{eqnarray}
with  $k=|\vk|$ and $q=|\vq|$ the momenta of the incoming photon 
and the outgoing meson respectively, $\sqrt {s} \equiv W$ the total energy of 
the system, $e^{- {({k}^2+{q}^2)}/{6\alpha^2_{ho}}}$ a form factor 
in the harmonic oscillator basis with the parameter $\alpha^2_{ho}$ 
related to the harmonic oscillator strength in the wave-function, 
and $M_{N^*}$ and $\Gamma(q)$ the mass and the total width of 
the resonance, respectively.  The amplitudes ${\mathcal A}_{N^*}$ 
are divided into two parts\cite{zpl97}: the contribution 
from each resonance below 2 GeV, the transition amplitudes of which 
have been translated into the standard CGLN amplitudes in the harmonic 
oscillator basis, and the contributions from the resonances above 2 GeV
treated as degenerate.

The contributions from each resonance 
is determined by introducing\cite{LS-1} a new set of 
parameters $C_{{N^*}}$, and the substitution rule 
${\mathcal A}_{N^*} \to C_{N^*} {\mathcal A}_{N^*}$, so that
${\mathcal M}_{N^*}^{exp} = C^2_{N^*}{\mathcal M}_{N^*}^{qm}$;
with ${\mathcal M}_{N^*}^{exp}$ the experimental value of 
the observable, and ${\mathcal M}_{N^*}^{qm}$ calculated in the 
quark model\cite{zpl97}. 
The $SU(6)\otimes O(3)$ symmetry predicts
$C_{N^*}$~=~0.0 for ${S_{11}(1650)} $, ${D_{13}(1700)}$, and 
${D_{15}(1675)} $ resonances, and $C_{N^*}$~=~1.0 for other
resonances in Table~2.  
Thus, the coefficients $C_{{N^*}}$ measure the discrepancies between 
the theoretical results and the experimental data and show the extent 
to which the $SU(6)\otimes O(3)$ symmetry is broken in the process 
investigated here.
%
%
\begin{table}[t]
\tbl{Nucleon resonances with $M \le$2 GeV and their 
assignments in $SU(6)\otimes O(3)$ configurations, masses, 
and widths.}
{\footnotesize
\begin{tabular}{llccccc}
\hline  
States & & $SU(6)\otimes O(3)$& & Mass & & Width   \\  
 & & & &  (GeV) & &  (GeV)  \\  \hline        
$S_{11}(1535)$&&$N(^2P_M)_{\frac 12^-}$&& && \\[1ex] 
$S_{11}(1650)$&&$N(^4P_M)_{\frac 12^-}$&&1.650&&0.150 \\[1ex]    
$D_{13}(1520)$&&$N(^2P_M)_{\frac 32^-}$&&1.520&&0.130\\[1ex]    
$D_{13}(1700)$&&$N(^4P_M)_{\frac 32^-}$&&1.700&&0.150\\[1ex]
$D_{15}(1675)$&&$N(^4P_M)_{\frac 52^-}$&&1.675&&0.150\\[1ex]
$P_{13}(1720)$&&$N(^2D_S)_{\frac 32^+}$&&1.720&&0.150\\[1ex]    
$F_{15}(1680)$&&$N(^2D_S)_{\frac 52^+}$&&1.680&&0.130\\[1ex]    
$P_{11}(1440)$&&$N(^2S^\prime_S)_{\frac 12^+}$&&1.440&&0.150\\[1ex]    
$P_{11}(1710)$&&$N(^2S_M)_{\frac 12^+}$&&1.710&&0.100\\[1ex]    
$P_{13}(1900)$&&$N(^2D_M)_{\frac 32^+}$&&1.900&&0.500\\[1ex]    
$F_{15}(2000)$&&$N(^2D_M)_{\frac 52^+}$&&2.000&&0.490\\[1ex] 
\hline
\end{tabular}\label{table2} }
\vspace*{-13pt}
\end{table}
%
%
One of the main reasons that the $SU(6)\otimes O(3)$ symmetry is
broken is due to the configuration mixings caused by the one-gluon
exchange\cite{IK}. 
Here, the most relevant configuration mixings are those of the
two $S_{11}$ and the two $D_{13}$ states around 1.5 to 1.7 GeV. The 
configuration mixings can be expressed in terms of the mixing angle
between the two $SU(6)\otimes O(3)$ states $|N(^2P_M)>$  and 
$|N(^4P_M)>$, with the total quark spin 1/2 and 3/2.
To show how the coefficients $C_{N^*}$ are related to the mixing angles, 
we express the amplitudes ${\mathcal A}_{N^*}$ in terms of the 
product of the photo and meson transition amplitudes
\begin{eqnarray}\label{eq:MixAR}
{\mathcal A}_{N^*} \propto <N|H_m| N^*><N^*|H_e|N>,
\end{eqnarray}
where $H_m$ and $H_e$ are the meson and photon transition operators,
respectively. For example, for the resonance ${S_{11}(1535)}$ 
Eq.~(\ref{eq:MixAR}) leads to
\begin{eqnarray}\label{eq:MixAS1}
{\mathcal A}_{S_{11}} \propto 
<N|H_m (\cos \theta _{S}
 |N(^2P_M)_{{\frac 12}^-}> - 
\sin \theta _{S}
|N(^4P_M)_{{\frac 12}^-}>) 
\nonumber\\ 
 (\cos \theta _{S} <N(^2P_M)_{{\frac 12}^-}| -
\sin \theta _{S} <N(^4P_M)_{{\frac 12}^-})|H_e|N>.
\end{eqnarray}
%
%
Then, the configuration mixing coefficients can be related to the
configuration mixing angles 
\begin{eqnarray}
C_{S_{11}(1535)} &=& \cos {\theta _{S}} ( \cos{\theta _{S}} - 
\sin{\theta _{S}}),\label{eq:MixS15} \\
C_{S_{11}(1650)} &=& -\sin {\theta _{S}} (\cos{\theta _{S}} + 
\sin{\theta _{S}}),\label{eq:MixS16} \\
C_{D_{13}(1520)} &=& \cos \theta _{D} (\cos\theta _{D} - 
\sqrt {1/10}
\sin\theta _{D}),\label{eq:MixD15} \\
C_{D_{13} (1700)} &=& \sin \theta _{D} (\sqrt {1/10}\cos\theta _{D} + 
 \sin\theta _{D}).\label{eq:MixD17}
\end{eqnarray}
%
%
\section{Results and Discussion}
\subsection{$\eta$-photoproduction channel}
we have fitted all $\approx$ 650 data points from recent measurements for 
both differential cross-sections\cite{Mainz,Graal,CLAS} and single 
polarization asymmetries\cite{pol}.
The adjustable parameters of our models are the $\eta NN$ coupling constants
and one $SU(6)\otimes O(3)$ symmetry
breaking strength coefficient ($C_{N^*}$) per resonance, except for the resonances
$S_{11}(1535)$ and $S_{11}(1650)$ on the one hand, and 
$D_{13}(1520)$ $D_{13} (1700)$ on the other hand, for which we introduce
the configuration mixing angles $\theta _{S}$ and $\theta _{D}$.
\begin{figure}[h]
\epsfysize=9.cm 
\centerline{\epsfxsize=4.7in\epsfbox{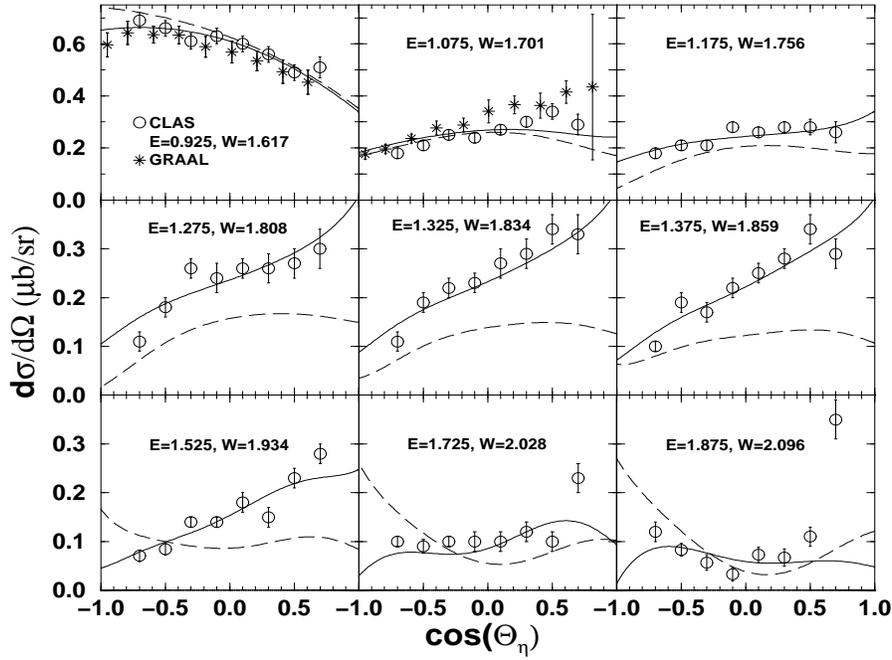}}   
\caption{Differential cross section for the process 
$\gamma p \to \eta p$: angular distribution at nine
incident photon energies ($E_{\gamma}^{lab}$), with the
corresponding total center-of-mass energy (W) also given;
units are in GeV.
The dashed curves are from the model embodying all known
three and four star resonances. The full curves show the
model including, in addition, a new $S_{11}$ resonance,
with M=1.780 GeV and  $\Gamma$=280 MeV.
CLAS (circles) and GRAAL (stars) data are
from Refs. [23] and [22], respectively.\label{inter}
}
\end{figure}
The first model includes explicitly all 
eleven known relevant resonances, mentioned above, with mass below 2 GeV, 
and the contributions from the known excited resonances above 2 GeV for a 
given parity. assumed to be degenerate and hence written in a 
compact form\cite{zpl97}.
In Fig.~1, we compare this model (dashed curves) to the data at nine
incident photon energies. As shown in our earlier works\cite{LS-2,LS-1},
such a model reproduces correctly the data at low energies
($E_\gamma ^{lab} \le$ 1 GeV). Above, the model misses the data. 
A possible reason for these theory/data discrepancies could be that
some yet unknown resonances contribute to the reaction mechanism.
We have investigated possible r\^ole played by extra $S_{11}$, $P_{11}$, and $P_{13}$
resonances, with three free parameters 
(namely the resonance mass, width, and strength) in each case.

By far, the most significant improvement was obtained by a third
$S_{11}$ resonance, with the extracted values M=1.780 GeV 
and $\Gamma$=280 MeV. The configuration mixing angles came out to be 
$\theta _{S}$=12$^\circ$ and $\theta _{D}$=-35$^\circ$, in agreement
with the Isgur-Karl model\cite{IK} and more recent predictions\cite{Karl}.
  
The outcome of this latter model is depicted in Fig.~1 (full curve)
and shows very reasonable agreement with the data, improving the reduced
$\chi^2$, on the complete data-base, by more than a factor of 2.
%
%
\subsection{Associated strangeness photoproduction channel}
The above formalism has also been used to investigate {\it all} 1640 recent data
points on the differential cross sections\cite{JLab,Saphir} for the
$\gamma p \to K^+ \Lambda$ reaction.
The adjustable parameters here are the $KYN$ coupling constant
and one $SU(6)\otimes O(3)$ symmetry
breaking strength coefficient ($C_{N^*}$) per nucleon resonance, as in the case
of the $\eta$-channel
(Table 2). Other nucleon resonances and all hyperon resonances in Table~1 
are included in a compact form\cite{zpl97} and bear no free parameters.

Figures 2 and 3 show our preliminary results for three excitation functions at
$\theta_{K}^{CM}$ = 31.79$^\circ$, 56.63$^\circ$, and 123.37$^\circ$ as a 
function of total center-of-mass energy ($W$). 
The choice of the angles is due to the published data by the JLab group\cite{JLab}.

Given the significant discrepencies between the two data sets, the minimization
procedure was performed as follows:

\begin{romanlist}
\item Both data sets were fitted {\it simultaneously}, leading to curve 
(a) in figures 2 and 3. 

\item Data sets from JLab\cite{JLab} and SAPHIR\cite{Saphir} 
were fitted {\it separately}. The 
curve (b) in Fig.~2 is obtained by fitting {\it only} the JLab data, while
the curve (d), Fig. 3, comes out from a fit {\it only} on the SAPHIR data.

\item The curves (a), (b), and (d) correspond to models embodying all
known resonances. At this stage, a third $S_{11}$ resonance was introduced,
in line with the $\eta$ case. With this additional resonance, the data from
JLab and SAPHIR were fitted seperately and the outcomes are the curves (c) 
and (e) in Figs. 2 and 3, respectively.

\end{romanlist}
%
\begin{figure}[t!]
\epsfysize=13.cm 
\centerline{\epsfxsize=4.7in\epsfbox{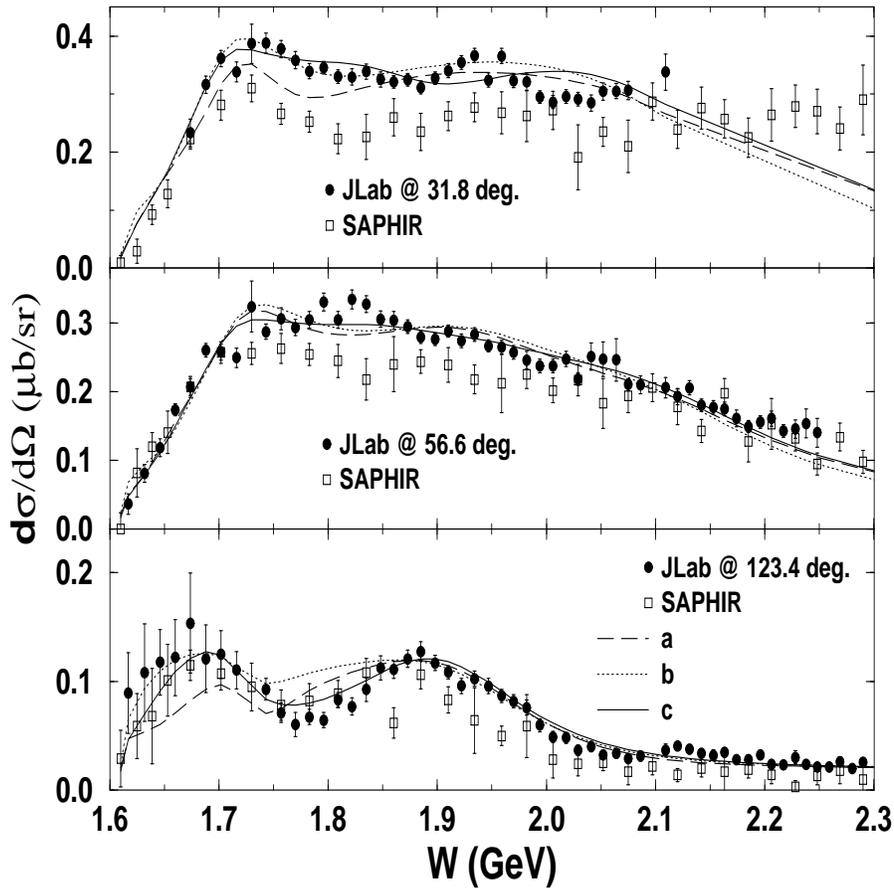}}   
\caption{Differential cross section for the process 
$\gamma p \to K^+ \Lambda$ as a function of total center-of-mass 
energy (W) in GeV. All the curves embody all known resonances.
The dashed curve (a) is from a fit to data from both JLab [25] and SAPHIR [26]. 
The dotted curve (b) is
obtained by fitting only JLab data.
The full curve (c) corresponds to this latter data set with an additional $S_{11}$
resonance. \label{Fig2}
}
\end{figure}
%
The model (a) gives a reduced $\chi^2$ of 3.7 (Table~3). Adding a third $S_{11}$ 
resonance, improves it slightly ($\chi^2$=3.5). However, fitting separately
each set of data, shows a significant sensitivity to the introduction of a
third $S_{11}$ resonance. Due to this latter new resonance,
for the JLab data the $\chi^2$ goes from 3.0 to 1.6, and for the SAPHIR data it
gets reduced from 2.1 to 1.4. We notice that in both cases, the SAPHIR data are
better reproduced within our approach.
\begin{figure}[t!]
\epsfysize=13.cm 
\centerline{\epsfxsize=4.7in\epsfbox{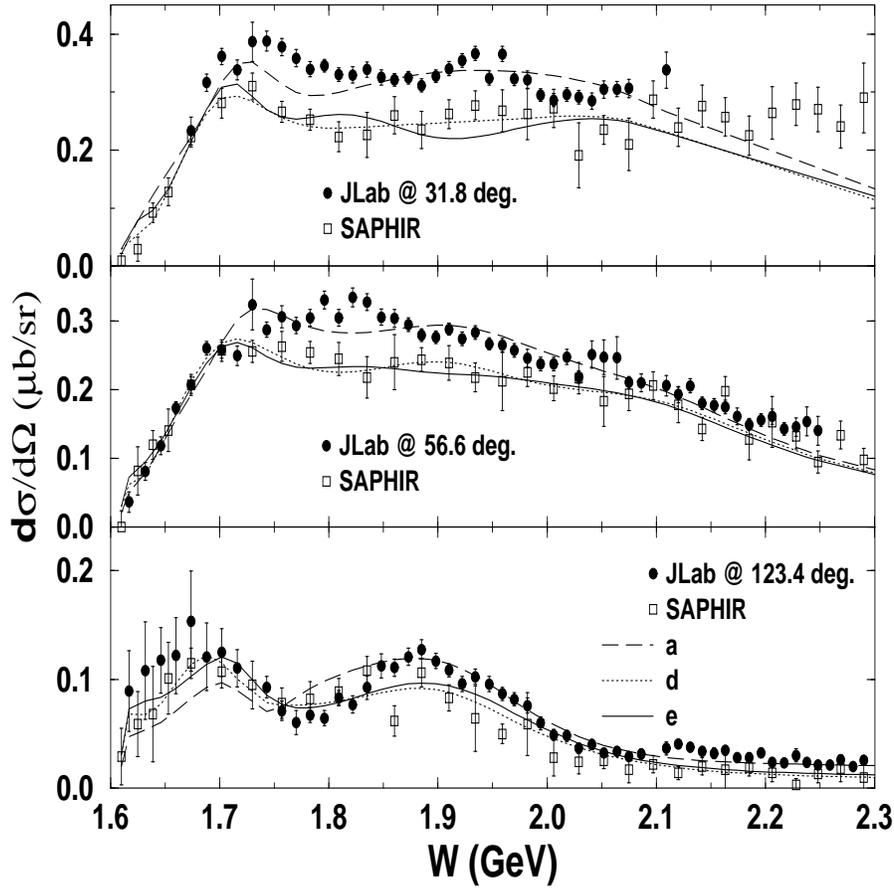}}   
\caption{Same as Fig. 2, except for curves (d) and (e). The dotted curve (d) is
obtained by fitting only SAPHIR data.
The full curve (e) corresponds to this latter data set with an additional $S_{11}$
resonance.\label{Fig3}
}
\end{figure}
%
%
\begin{table}[t!]
\tbl{Summary of models (a) to (e).}
{\footnotesize
\begin{tabular}{ccccc}
\hline
 Model &  Data  &  \# of data points & Reduced $\chi^2$ & 3$^{rd}~ S_{11}$ \\
\hline
 a & JLab \& SAPHIR  & 1640  &  3.7 & \\
 b & JLab            &  920  &  3.0 & \\
 c & JLab            &  920  &  1.6 & $M$=1.852 GeV ; $\Gamma$=187 MeV \\
 d & SAPHIR          &  720  &  2.1 & \\
 e & SAPHIR          &  720  &  1.4 & $M$=1.835 GeV ; $\Gamma$=246 MeV \\
\hline
\end{tabular}\label{table3} }
%
\end{table}
%
%
 
\subsection{New $S_{11}$ resonance}
Several authors\cite{LS-2,LS-3,LW96,Zagreb,BES,Gia,chen,Try}
have reported on a third $S_{11}$ resonance with a mass around 1.8 GeV 
(see Table~4).
Our chiral constituent quark approach applied to the
$\gamma p \to \eta p, K^+ \Lambda$ reactions puts the mass in the range of
1.780 to 1.852 GeV and the width between 187 and 280 MeV. This dispersion is,
at least partly, due to the discrepancies among data reported by different
collaborations.
The extracted values for the mass and width from the $\gamma p \to \eta p$
process are consistent with those predicted by the authors of Ref.\cite{LW96}
(M=1.712 GeV and $\Gamma$=184 MeV), 
and our previous findings\cite{LS-2}.
Moreover, for the one star $S_{11}(2090)$ resonance\cite{PDG}, the 
Zagreb group $\pi N$ and $\eta N$
coupled channel analysis\cite{Zagreb} produces the following values
M = 1.792 $\pm$ 0.023 GeV and $\Gamma$ = 360 $\pm$ 49 MeV.
The BES Collaboration
reported\cite{BES} on the measurements of the
$J/\psi \to p \overline{p} \eta$ decay channel. 
In the latter work, a partial wave analysis
leads to the extraction of the mass and width of the 
$S_{11}(1535)$ and $S_{11}(1650)$ resonances, and the authors find  
indications for an extra resonance with 
M = 1.800 $\pm$ 0.040 GeV, and $\Gamma$ = 165$^{+165}_{-85}$ MeV.
A more recent work\cite{Gia} based on the hypercentral constituent 
quark model, 
predicts a missing $S_{11}$ resonance with M=1.861 GeV.
Finally, a self-consistent analysis of pion scattering and photoproduction
within a coupled channel formalism, iconcludes\cite{chen} on the 
existence of a third $S_{11}$ resonance with M =1.803 $\pm$ 0.007 GeV.
%
%
\begin{table}[h!]
\tbl{Summary of studies on a 3$^{rd}$ $S_{11}$ resonance.}
{\footnotesize
\begin{tabular}{cclc}
\hline
 Mass &  Width &  Comment & Ref. \\
 (GeV)&  (MeV) &   &  \\
\hline
 1.780 & 280  &  CQM applied to $\gamma p \to \eta p$ & 28; Sec. 3.1 \\
 1.835 & 246  &  CQM, applied to $\gamma p \to K^+ \Lambda$ data from SAPHIR & Sec. 3.2 \\
 1.852 & 187  &  CQM, applied to $\gamma p \to K^+ \Lambda$ data from JLab & Sec. 3.2 \\
 1.730 & 180  & $KY$ molecule  & 29 \\
 1.792 & 360  & $\pi N$ and $\eta N$ coupled-channel analysis  &  30 \\
 1.800 & 165  & $J/\Psi$ decay & 31 \\
 1.861 &   & Hypercentral CQM & 32 \\
 1.846 &   & Pion photoproduction coupled-channel analysis  & 33 \\
\hline
\end{tabular}\label{table4} }
\vspace*{-13pt}
\end{table}
%
%
\section{Summary and Concluding remarks}
In this contribution, the results of a chiral constituent 
quark model have been compared with the most recent published data 
on the $\gamma p \to \eta p, K^+ \Lambda$ processes, with emphasize on
a third $S_{11}$ resonance. Our results are consistent with findings
by other authors\cite{LW96,Zagreb,BES,Gia,chen}, showing evidence for 
such a new resonance, with $M \approx$1.8 GeV and $\Gamma \approx$250 MeV.

In the case of the $\eta$ photoproduction channel, we need to extend
our analysis to the very recent data from ELSA\cite{ELSA}. 

The associated strangeness production channel suffers at the present time
from discrepancies between the two copious data set from JLab and SAPHIR. 
New data from GRAAL will hopefully provide us with a better vision of the
experimental situation. Within our approach, a more comprehensive interpretation
of the $K^+ \Lambda$ channel is underway with respect to the polarization
observables\cite{LEPS}, as well as to the observables\cite{JLab,Saphir} of the 
$\gamma p \to K^+ \Sigma^\circ$.

To go further, the coupled channel effects\cite{CC1,CC2,CC3,CC4,CC5,CC6} 
have to be considered. The effect of the multi-step process 
$\gamma p \to \pi N \to K^+ \Lambda$  has been reported\cite{CC4} to 
be significant at the level of inducing 20\% changes on the total
cross section of
the direct channel ($\gamma p \to K^+ \Lambda$).
The dynamics of the intermediate states $\pi N \to KY$, as well as final
states interactions $KY \to KY$, with $Y \equiv \Lambda, \Sigma$ have
been recently studied\cite{CC5} within a dynamical coupled-channel model of
meson-baryon interactions. Those efforts deserve to be extended to the 
multi-step processes 
$$\gamma p \to \pi N \to \eta p~,~K^+ \Lambda~,~K^+ \Sigma ^\circ~,~K^\circ \Sigma ^+,$$
and also embody the final state interactions.

\bigskip

It is a pleasure for me to thank K.H. Glander and R. Schumacher for 
having provided me with the complete set of SAPHIR and JLab data, respectively, 
prior to publication. I am indebted to my collaborators W.T. Chiang, C. Fayard, 
T.-S. H. Lee, Z. Li, T. Mizutani, and F. Tabakin.

\end{document}